\documentclass[%
 reprint,
superscriptaddress,
 amsmath,amssymb,
aps,
prl,
]{revtex4-1}
\usepackage{graphicx}
\usepackage{units}
\usepackage{tabularx,booktabs}
\usepackage{datetime}
\usepackage{color}
\usepackage[normalem]{ulem}
\usepackage{mathtools}

\renewcommand{\Im}{{\rm Im}}

\renewcommand{\d}{{\rm d}}

\newcommand{\per}{{\rm per}}

\newcommand{\opt}{{\rm opt}}
\newcommand{\inj}{{\rm inj}}

\newcommand{\dBm}{{\,\rm dBm}}
\newcommand{\ns}{{\rm ns}}

\newcommand{\GHz}{{\rm GHz}}
\newcommand{\mA}{{\rm mA}}

\begin{document}

\title{Devil's Staircases in Continuous Systems with Modulated Forcing}

\newcommand{\UCC}{Department of Physics, University College Cork, Ireland}
\newcommand{\Tyndall}{Tyndall National Institute, Cork, Ireland.}

\author{Benjamin Lingnau}  \affiliation{\UCC} \affiliation{\Tyndall}
\email{benjamin.lingnau@ucc.ie}
\author{Kevin Shortiss} \affiliation{\UCC} \affiliation{\Tyndall}
\author{Fabien Dubois}  \affiliation{\UCC} \affiliation{\Tyndall} 
\author{Frank H. Peters}  \affiliation{\UCC} \affiliation{\Tyndall} 
\author{Bryan Kelleher}  \affiliation{\UCC} \affiliation{\Tyndall} 
\date{\today, \currenttime}

\begin{abstract}
The discrete circle map is the archetypical example of a driven periodic system, showing a complex resonance structure under a change of the forcing frequency known as the devil's staircase. Adler's equation can be seen as the direct continuous equivalent of the circle map, describing locking effects in periodic systems with continuous forcing. This type of locking produces a single fundamental resonance tongue without higher order resonances, and a devil's staircase is not observed.
%
%
We show that, with harmonically modulated forcing, nonlinear oscillations close to a Hopf bifurcation generically reproduce the devil's staircase even in the continuous case. Experimental results on a semiconductor laser driven by a modulated optical signal show excellent agreement with our theoretical predictions. The locking appears as a modulation of the oscillation amplitude as well as the angular oscillation frequency.
%
Our results show that by proper implementation of an external drive, additional regions of stable frequency locking can be introduced in systems which originally show only a single Adler-type resonance tongue.
\end{abstract}


\maketitle


The synchronization of periodic systems by an external force or coupling to another oscillator has been the topic of intensive research {for several} centuries \cite{PEC90,STR94a,ROS96,PEC97,PIK97b,BOC02,PIK03,ARE08,BAL09,BAE12}. The entrainment of the intrinsic oscillation frequency of an oscillator by a driving signal has been observed in a multitude of dynamical systems, from electronic circuits \cite{ADL73,CHU93}, {through} climate dynamics \cite{BEN82,CRU12}, to physiological \cite{STR93,GLA01,GOL10b} and neuronal systems \cite{HAN95c,DIE99}.
%
When the detuning $\nu$ between the driving frequency and {an} oscillator's intrinsic
frequency is sufficiently small, the oscillation frequency of the driven system becomes entrained by the external {signal}. Often, outside of this main locking region, no further resonances exist. This is usually the case when the intrinsic oscillation frequency is very large compared to the width of the resonance region, and 
higher harmonic resonances are unsupported by the system. We refer to this type of locking as Adler-type, as such systems can be approximated by Adler's equation:
\begin{align}
  \frac{\d}{\d t} \phi(t) = -\nu - K \sin\phi(t)\,.  \label{eq:adler}
\end{align}
This equation describes phase-locking in a continuously driven oscillator close to resonance \cite{ADL73,PIK03}. 
Adler's model exhibits phase-locking for sufficiently strong driving strength $K \geq |\nu|$, in which the driven oscillator maintains a constant phase difference $\phi$ {relative} to the external force. Beyond this, the only remaining solution is the drifting-phase solution, in which the {amplitude of the} phase difference {monotonically} increases in time and the average oscillation frequency approaches its free-running value for $|\nu|\to\infty$.
\begin{align}
  \phi_{n+1} = \phi_n - \tilde\nu - k \sin\phi_n\,, \label{eq:circlemap}
\end{align}
In contrast to the continuous forcing of the Adler model, the circle map describes the dynamics of periodic systems with {pulsed} forcing, with a multitude of applications including neuronal \cite{HAN95b,COO99} and biological systems \cite{MIR90a,ERM91}. 
The resulting pattern of locking is referred to as Arnold-type and contrasts strongly with the single resonance region obtained in the continuous case. In addition to the main resonance tongue, additional Arnold tongues appear at oscillation frequencies corresponding to rational ratios of the driving frequency $T^{-1}$, leading to the formation of a devil's staircase \cite{JEN83}. Such a devil's staircase has been observed in a multitude of different physical and mathematical systems \cite{BRO97b}, including chemical oscillations \cite{MAR94a}, interfaces between crystalline solids \cite{BAK82}, and particles in periodic potentials \cite{WIE01a}. 
The richer synchronization dynamics of the circle map compared to Adler's equation stem from the instantaneous driving term and the related discrete phase steps that allow the system trajectory to pass the original fixed points of Eq.~\eqref{eq:adler} without travelling through the points. This leads to the emergence not only of a devil's staircase locking structure, but also quasiperiodic and chaotic dynamics \cite{ZAS78,GLA94a}.
%
Naively, one might reason that a periodic modulation of the driving frequency would induce Arnold-type locking. In fact, this is not sufficient for physical systems and an additional nonlinearity in the system is required. We introduce this by moving beyond a pure phase oscillator.

In this Rapid Communication, we thus study the emergence of Arnold-type locking in the oscillation frequency and amplitude in continuous systems which show only Adler-type locking around a single main resonance tongue for continuous forcing. We show that our result is universal and that the devil's staircase structure emerges generically in all nonlinear oscillators close to Hopf bifurcations with periodically modulated forcing. As an illustration of the phenomenon, we investigate a semiconductor laser with modulated optical forcing in experiment and theory, revealing the possibility to induce tailored devil's staircases by a proper choice of the driving parameters.


\begin{figure}
 \includegraphics[width=\columnwidth]{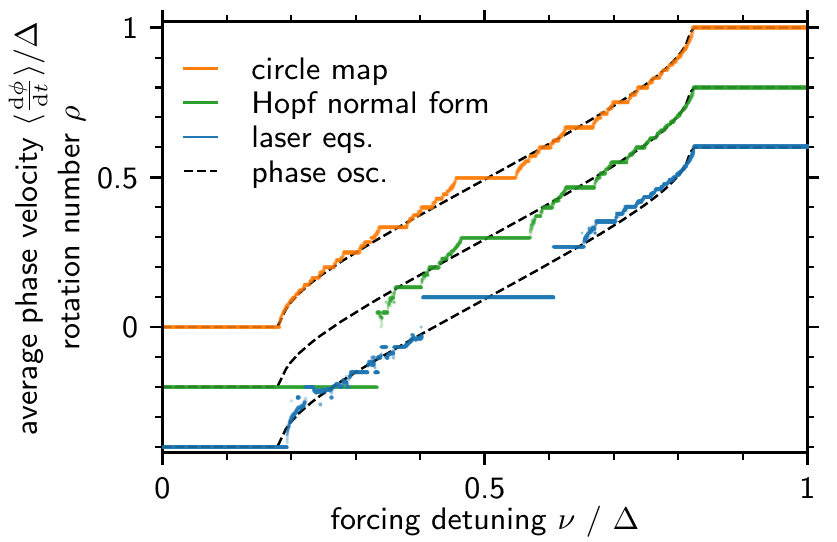}
 \caption{Comparison of the locking behaviour of different driven systems, showing the average phase velocity for the circle map \eqref{eq:circlemap} (orange, $K=0.18$), the Hopf normal form \eqref{eq:hopf_normal_form} (green, $K=0.46$, $\gamma=1$, $\Delta=0.5$), and the laser equations (blue, $K=0.06$, $\beta=0$). We plot subsequent curves with an artificial vertical offset of $0.2$ for better readability.
 The driven phase oscillator \eqref{eq:continuousEquivalent} (dashed line, $K=0.18$) is shown as comparison for each model.
 }
 \label{fig:lockingcomparison}
\end{figure}

The circle map Eq.~\eqref{eq:circlemap} can be written as a one-dimensional ordinary differential equation by expressing the discrete periodic forcing as a temporal Dirac comb, acting at times $n T$. The values $\phi_n$ of the circle map are thus obtained by evaluating the continuous variable $\phi(t)$ at the times of the forcing, $\phi_n := \phi(n T)$. The dynamical equation for $\phi(t)$ is then given by:
\begin{align}
  \frac{\d}{\d t} \phi(t) 
  &= -\frac{\tilde\nu}{T} - k \sin\phi(t) \times \sum_{\mathclap{n=-\infty}}^\infty \delta(t - n T) \notag\\
  &= -\frac{\tilde\nu}{T} - \frac{\Delta}{2\pi} k \sin\phi(t) \times \sum_{\mathclap{n=-\infty}}^\infty \exp\left( i n \Delta\, t \right). \label{eq:continuousEquivalent}
\end{align}
The discrete, periodic forcing in time-discrete maps can be understood as a forcing of the continuous system with an infinite number of driving frequencies, each spectrally separated by $\Delta := 2\pi/T$. By identifying $\frac{\tilde\nu}{T}=\nu$ and $\frac{\Delta}{2\pi} k = K$, the circle map is formally equivalent to Adler's equation with an infinite number of spectrally equidistant driving terms.
%
In the limit of an infinite number of spectral components the forcing introduces periodic discontinuities in the trajectory $\phi(t)$, leading to the formation of Arnold-type locking.
When the forcing is instead restricted to any finite number of Fourier components as is inevitably the case in real-world examples, the resulting solution of $\phi(t)$ will remain continuous \cite{ENG12a}. Thus, even though the explicit time dependence of the driving term will increase the dimensionality of the system's phase space, the phase oscillator with a periodic driving will show only Adler-type locking around each of the spectral constituents of the forcing term. This is demonstrated in Fig.~\ref{fig:lockingcomparison} (top lines), where we compare the locking behavior of the circle map and the continuous phase oscillator with a finite number of forcing terms by restricting the sum to the three central terms $n\in\{-1,0,1\}$. We evaluate the average phase velocity $\langle\frac{\d \phi}{\d t}\rangle$ or, in the case of the circle map, the rotation number $\rho:=\lim\limits_{n\to\infty}\frac{\phi_n}{n}$ as a function of the detuning $\nu$, swept between two adjacent main resonances.
%
%
%
The circle map shows Arnold-type locking. There is a devil's staircase between the main resonances with increasingly smaller regions of subharmonic locking centered at fractional ratios of the forcing frequency, i.e. $\rho=\frac pq \Delta$ with $p,q\in\mathbb N$. The continuous phase oscillator model with three forcing components clearly shows the main resonances (near $\nu/\Delta = 0~ \text{and} ~1$) but no devil's staircase is observable. Instead, only a smooth transition of the average phase velocity in between these resonances can be observed. Its locking is thus of Adler type, with the main resonance tongues but no further subharmonic resonances.
We will see, however, that the combination of a finite number of injected Fourier components and intrinsic nonlinearities results in the emergence of a devil's staircase structure with an Arnold-type locking.


\begin{figure*}[bt!]
  \includegraphics[width=0.6\columnwidth]{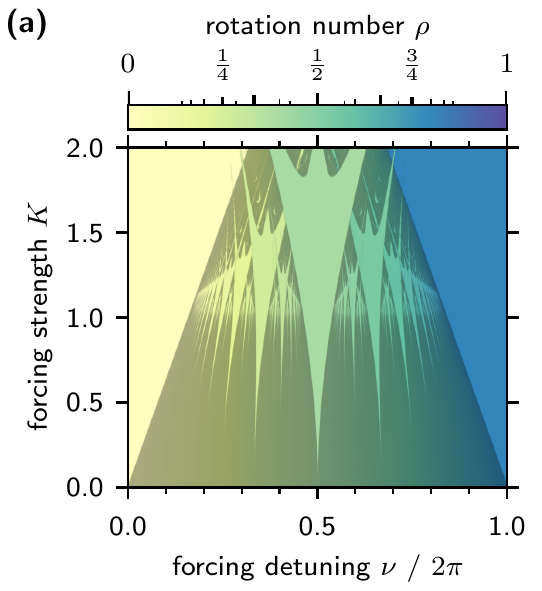}
  \includegraphics[width=0.709\columnwidth]{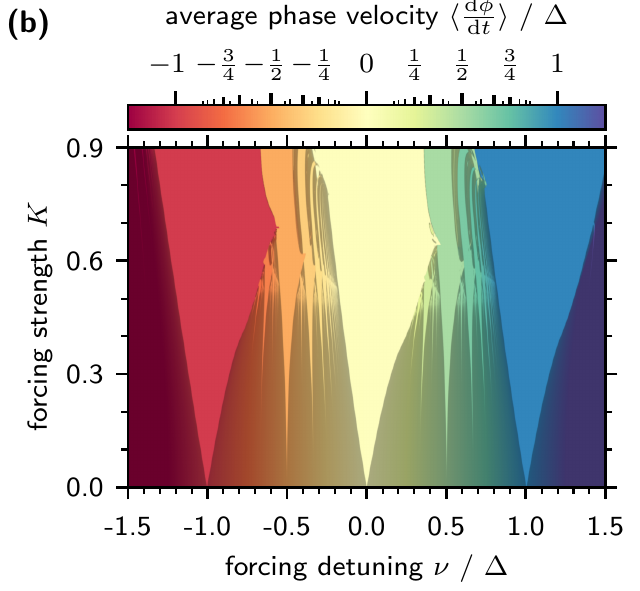}
  \includegraphics[width=0.709\columnwidth]{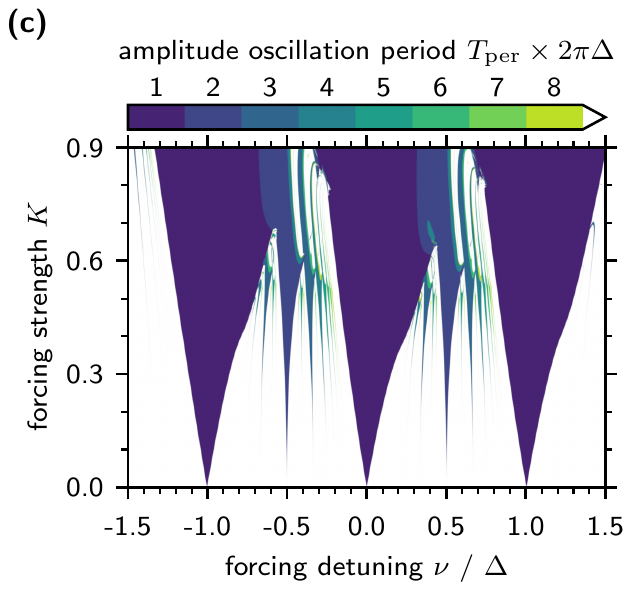}
 \caption{Two-dimensional resonance diagrams. \textbf{(a)}~Arnold tongues in the circle map, and \textbf{(b)}~in the HNF oscillator with modulated forcing. \textbf{(c)}~Oscillation period of the amplitude oscillations in the HNF oscillator. The maps are shown in dependence of the injection strength $K$ and the driving frequency $\nu$.
 The plots in {(a)}, {(b)} highlight areas where the rotation number changes little between neighboring sampling points, i.e., $|\frac{\d\rho}{\d\nu}| < 0.05$ and $|\frac{\d}{\d\nu}\langle\frac{\d\phi}{\d t}\rangle| < 0.05$, respectively. The HNF is simulated with $\gamma=1$, $\Delta=0.5$.
 }
 \label{fig:2dbifHopf}
\end{figure*}

We consider a Hopf-normal-form (HNF) oscillator describing the periodic orbit born in a supercritical Hopf bifurcation in all nonlinear systems \cite{GAR12b}. The ubiquity of such orbits in nonlinear dynamical systems means that this is a particularly important case. As with Adler's equation before, we additionally introduce a periodically modulated forcing:
\begin{align}
  \frac{\d z}{\d t} &= \Big[ \big(1\!-\!|z|^2\big)(1\!+\!i\gamma) - i\nu \Big]z + K \big( 1 \!+\! 2 \cos(\Delta\, t) \big) . \label{eq:hopf_normal_form}   
\end{align}
Here, $z$ describes the complex normalized oscillation amplitude, and $\gamma$ is the shear parameter, which introduces a coupling between the orbit's amplitude and oscillation frequency. We introduce a modulated forcing via the forcing strength $K$, the forcing frequency $\nu$ relative to the intrinsic oscillation frequency, and the modulation frequency $\Delta$.
We repeat the investigation of the locking behavior for the HNF oscillator by evaluating the average oscillator frequency $\langle\frac{\d\phi}{\d t}\rangle = -\langle\Im\big[\frac{\d z/\d t}{z(t)}\big]\rangle$, shown in Fig.~\ref{fig:lockingcomparison} (middle lines). With the same type of modulated forcing as in the phase oscillator before, the HNF can be clearly seen to reproduce Arnold-type locking. For $\nu<\Delta/2$, the shear parameter in the time-continuous models leads to a slight asymmetry in the dynamics and a breakup of the staircase. The Hopf normal form thus forms a ``harmless'' staircase \cite{BAK82}, with discontinuous jumps. We note that for different choice of parameters, even without shear ($\gamma=0$ in Eq.~\eqref{eq:hopf_normal_form}), the devil's staircase structure persists, albeit less pronounced. The locking plateaus visible in the average phase velocity of the driven HNF not only correspond to a locking of the oscillator frequency but are accompanied by a locking of the oscillation amplitude. 
The amplitude $|z|$ is modulated by the time-varying forcing strength and performs oscillations with a period $T_\per = q T$, equal to an integer multiple $q$ of the forcing period.

To investigate the global locking structure of the forced oscillators, we calculate two-dimensional bifurcation diagrams in the detuning $\nu$ and the injection strength $K$ in Fig.~\ref{fig:2dbifHopf}. For comparison, we reproduce the Arnold tongue structure of the circle map in Fig.~\ref{fig:2dbifHopf}(a). For the continuously driven HNF oscillator we evaluate the average phase velocity in Fig.~\ref{fig:2dbifHopf}(b), as well as the oscillation period $T_\per$ of the amplitude $|z(t)|$ in Fig.~\ref{fig:2dbifHopf}(c).
We highlight areas where the frequency changes little with the detuning, i.e., where it is locked. Both the amplitude oscillation period and the frequency show pronounced Arnold tongues around the three comb lines, with higher harmonic locking tongues for detuning frequencies in between. The locking boundaries in the frequency and amplitude coincide. The comparison with the circle map shows the striking correspondence between the two model systems.
The shear parameter $\gamma$ lifts the symmetry of the driven system with respect to the sign of the detuning $\nu$, introducing an asymmetry in the shape of the locking tongues as seen earlier in Fig.~\ref{fig:lockingcomparison}. The locking boundaries are thus shifted slightly towards positive $\nu$ for stronger forcing $K$.




\begin{figure*}
\includegraphics[width=.9\columnwidth]{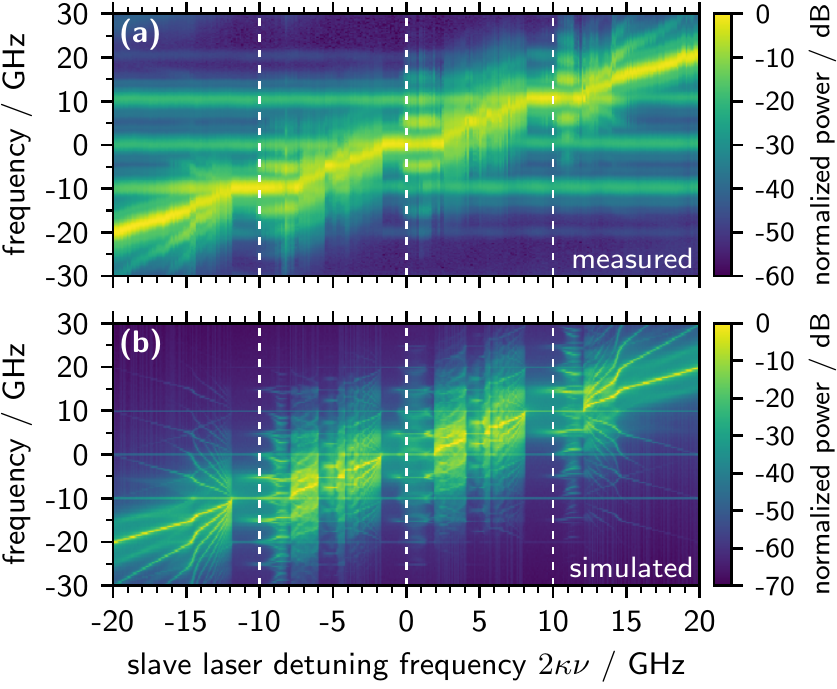}
\includegraphics[width=.9\columnwidth]{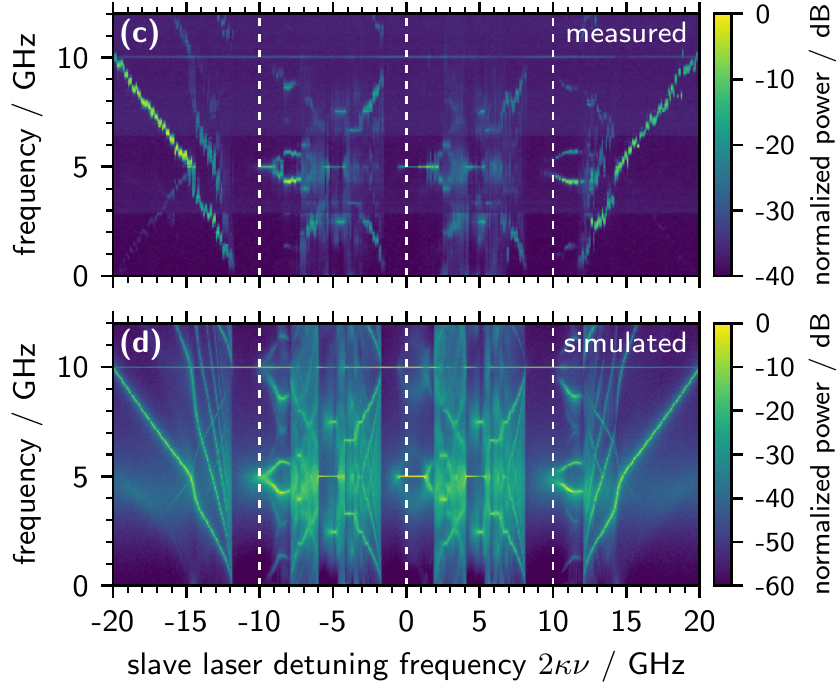}
 \caption{\textbf{(a,b)} Experimental and simulated optical spectra, and \textbf{(c,d)} power spectra of the laser output for a sweep of the relative frequency detuning $\nu$ of the slave laser. The injected signal consists of three spectral lines with a spacing $2\kappa\Delta=2\pi \times 10~\GHz$. The experiments were performed at a slave laser pump current of $90\mA$ ($1.75$ times its threshold).
The injected optical power is $-5\dBm$, simulated with $K=0.06$. The vertical dashed white lines show 
where the free-running laser frequency is resonant to one of the comb lines.}
\label{fig:linescan10GHz}
\end{figure*}


To illustrate our results with a real-world example, we analyse the dynamics of a semiconductor laser subject to a periodically modulated external optical input signal. Laser systems have provided fertile testbeds for the study of driven nonlinear oscillations \cite{SOR13}, readily displaying a variety of different dynamical effects, including synchronization between coupled lasers \cite{HOH97a,FIS00,BON12,CLE14}, and locking to an external signal \cite{VAR97} among many others. 
Devil's staircases and related phenomena have also been observed in laser systems and, in particular, in pulsing lasers with external current modulation \cite{BAU89} and lasers undergoing feedback from an external cavity \cite{TYK16,JAU17}. The nonlinear dynamics of lasers with a modulated or pulsing optical forcing has recently gained increasing attention \cite{SHO19, DOU20}. In \cite{LIN09b} a devil's staircase was numerically obtained via analysis of a laser subject to a repetitive forcing from strong and short optical pulses.
In laser systems driven by a constant optical field the simple harmonic coupling term $\sin\phi$ as in Eq.~\eqref{eq:adler} arises naturally \cite{TAR95a}.
In fact, Adler's equation is the low injection limit of the injected laser system \cite{ERN10b,ZIE13}. The laser system itself is nonlinear, which leads to rich dynamic scenarios in optically injected lasers, including high-order periodic, quasiperiodic, and chaotic dynamics \cite{NIZ99,WIE05}. However, in the weak continuous wave injection regime, the locking to the driving signal is of Adler-type, comprising a single resonance tongue. 

In the case of optically injected lasers, the Fourier sum in Eq.~\eqref{eq:continuousEquivalent} corresponds to the injection of an optical frequency comb, or, equivalently, a periodically modulated continuous wave signal.
We thus employ a single-mode master laser 
and modulate its output using a Mach-Zehnder modulator driven by a $10~\GHz$ sinusoidal signal. 
The resulting optical signal consists of three optical lines of nearly equal optical power, separated by the modulation frequency, which is then injected into a single-mode semiconductor laser. 
The driving signal corresponds to taking the three center terms of the Dirac comb with $n\in\{-1,0,1\}$ in Eq.~\eqref{eq:continuousEquivalent}. We control the frequency detuning between the slave laser and the injection signal via the laser mount temperature, which shifts the frequency of the slave laser cavity mode. 
We record the power spectra and optical spectra of the laser output under a continuous sweep of the slave laser detuning,  shown in Fig.~\ref{fig:linescan10GHz}. 
The optical spectra in Fig.~\ref{fig:linescan10GHz}(a) show the three comb lines with a spacing of $10~\GHz$ and the slave laser frequency, swept from {a detuning of} $-20~\GHz$ to $20~\GHz$. When the laser is tuned close to one of the three comb lines (vertical dashed lines in Fig.~\ref{fig:linescan10GHz}), its emission frequency becomes locked to that comb line. At the right edge within these locking regions, more complex features become apparent, with additional spectral components in between the comb lines, but {with} the dominant frequency component remaining locked to the comb. 
Between the main locking regions, we observe less pronounced locking regions to frequencies in between two adjacent comb lines. As the resolution of the optical spectrum analyser limits the visibility of these resonances, we additionally record the corresponding power spectra of the laser output, shown in Fig.~\ref{fig:linescan10GHz}(c).
Across the whole detuning range, a strong spectral component at 10~$\GHz$ is visible, corresponding to the beating frequency between the injected comb lines. When the laser is locked {to the individual comb lines}, this beating frequency and its harmonics (not shown) are the only features in the spectrum. Increasing the detuning from these regions, a subharmonic locking to half the comb spacing at $5~\GHz$ can be seen for the two leftmost locking regions. Subsequently, an unlocking and splitting of the signal at $5~\GHz$ can be observed (e.g., between $[-10~\GHz,-8~\GHz]$). In these regions, the optical spectrum, however, still suggests a frequency locking of the laser to the respective comb line. In between the locking regions, additional spectral lines appear, increasing from zero frequency to $10~\GHz$ and vice versa, while sweeping the detuning between the edges of adjacent locking regions. Within this unlocked detuning interval, quasiperiodic and chaotic dynamics can be observed, with intermittent harmonic locking to rational fractions of the comb frequency spacing. This structure in the power spectrum of the laser output closely resembles a devil's staircase structure, with prominent spectral signatures at $5~\GHz$, $3.3~\GHz$, and $2.5~\GHz$, corresponding to $\tfrac12$, $\tfrac13$, and $\tfrac14$ of the comb spacing. For very large detuning beyond the locking regions of the outermost comb lines, no additional higher order resonances can be observed. 
%


In order to perform a deeper analysis of the involved locking dynamics, we formulate a dimensionless rate equation model \cite{ERN10b} with an injected optical signal, modelling the dynamics of the complex electric field inside the cavity, $E(t)$, and the normalized optical gain $N(t)$:
\begin{align}
 \frac{\d}{\d t} E(t) &= (1+ i \alpha) \frac{N(t)}{2} E(t) + \frac{\partial E}{\partial t }  \Big|_{\inj} + \sqrt{\beta} \ \xi (t) \\
 T \frac{\d}{\d t}  N(t) &= J - N(t) - (N(t)+1) |E(t)|^2
\end{align}
Here, $J=1.5$ is the normalized pump current, $T = 26.4$ is the relative inversion lifetime, and $\alpha = 3$ is the amplitude-phase coupling parameter. The time $t$ is given in units of the inverse optical cavity lifetime, $2\kappa = 120~\ns^{-1}$.
%
%
The optical comb injected into the laser cavity is modeled by an additional driving term:
\begin{align}\label{eq:injection}
 \frac{\partial E}{\partial t} \Big|_{\inj} &= \frac{K}{2} E_0 \big[ 1 + 2 m \cos(\Delta\, t) \big] -  i \nu E(t)\,, 
\end{align}
where $m = 1.1$ is the relative strength of the injected side-modes at frequencies $\pm\Delta$. 
The injection strength $K$ is the amplitude ratio of the injected field and the free-running laser intracavity field $E_0=\sqrt{J}$. The final term in Eq.~\eqref{eq:injection} transforms the electric field into the rotating frame of the central comb mode, at a frequency detuning of $\nu$ with respect to the free-running laser frequency \cite{WIE05}. 
We include the spontaneous emission noise inside the laser cavity by a $\delta$-correlated complex Gaussian white noise source term $\xi(t)$, with a noise strength $\beta = 4\times 10^{-5}$ .

\begin{figure}[t]
  \includegraphics[width=0.9\columnwidth]{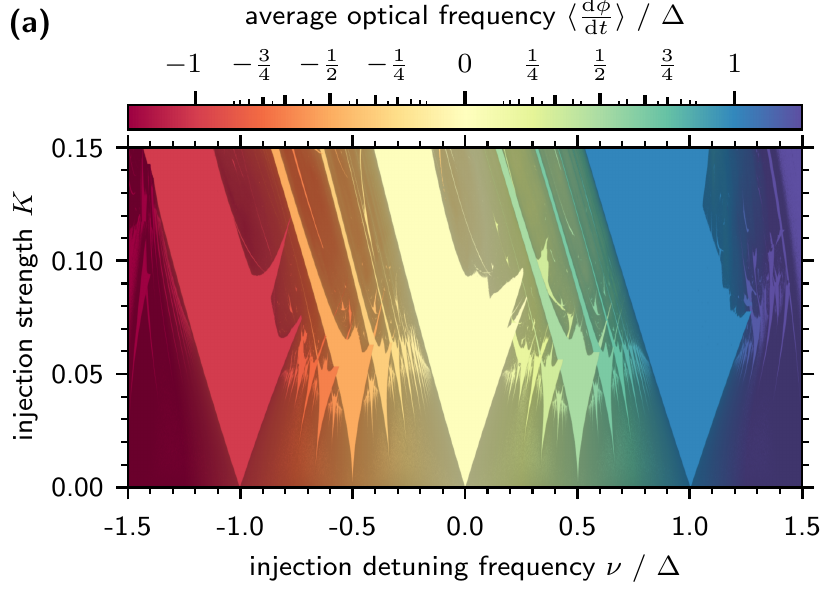}\\
  \includegraphics[width=0.9\columnwidth]{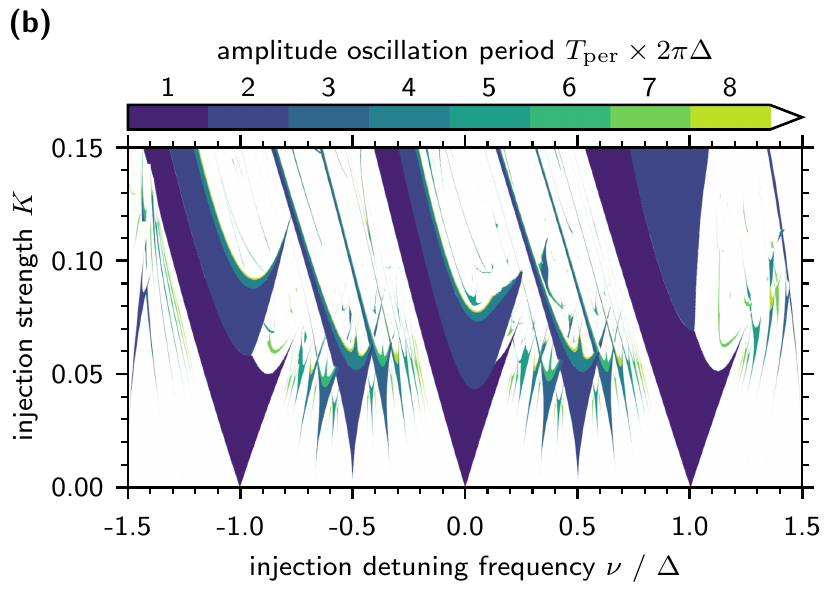}
 \caption{Two-dimensional resonance diagrams for the semiconductor laser model, showing \textbf{(a)} the average optical frequency $\langle\nu_\opt\rangle$, and \textbf{(b)} the period $T_\per$ of the intensity oscillations, depending on the injection strength $K$ and the laser detuning $\nu$ from the center comb line with respect to the free-running slave laser frequency, cf. Fig.~\ref{fig:2dbifHopf}.
Simulated without spontaneous emission noise, $\beta=0$.
All frequencies are normalized to the comb spacing $\Delta$. 
 }
 \label{fig:2dbif}
\end{figure}

The numerically calculated optical and power spectra of the laser emission under the external comb injection are shown in Fig.~\ref{fig:linescan10GHz}(b) and (d). The model closely reproduces the measured dynamics and locking behaviour, with a clear devil's staircase structure appearing between the comb lines. In Fig.~\ref{fig:lockingcomparison} (bottom lines) we show the dependence of the average lasing frequency, $\langle\nu_\opt\rangle := \langle{\d\phi}/{\d t}\rangle$, on the detuning for the laser equations, reproducing the devil's staircase in the circle map very closely. In the experimental optical spectra, $\langle\nu_\opt\rangle$ denotes the position of the dominating spectral line of the laser emission.
%
%
%
Two-dimensional bifurcation diagrams in the detuning $\nu$ and the injection strength $K$, shown in Fig.~\ref{fig:2dbif}, reveal the full devil's staircases in the optically injected semiconductor laser.
Both the lasing frequency and the amplitude oscillation period show pronounced Arnold tongues around the three comb lines, with higher harmonic locking tongues for detuning frequencies in between. In contrast to continuous wave injection, exact phase locking with a constant phase difference between the slave laser and injection signal is impossible in our setup, due to the time-varying injected signal. The frequency locking that we observe is therefore always a locking of the average optical frequency, while the phase is non-static. In lasers with single-mode injection a similar type of locking exists close to Hopf bifurcations \cite{THE11,KEL12,THO16}. 
The locking tongues in Fig.~\ref{fig:2dbif} exhibit a pronounced asymmetry due to the $\alpha$-parameter, which is equivalent to the shear parameter in the HNF oscillator. In semiconductor lasers the value of $\alpha$ can be quite large,
which reduces the forcing strength at which the harmonic locking tongues appear while also increasing their relative extent in parameter space. Fig.~\ref{fig:2dbif}(b) reveals a break-up and period doubling of the fundamental Arnold tongues at higher $K$, which is not observed for the HNF oscillator and is induced by the additional nonlinearities of the laser system \cite{SHO19}.
The newly generated harmonic locking regions between the fundamental locking tongues can be tuned by choosing the modulation frequency $\Delta$ appropriately, allowing the locking to a desired harmonic resonance tongue in between comb lines. Our earlier analysis demonstrates that this effect is not limited to the control of lasers, but is in fact universal and can be applied to any nonlinear oscillator in the vicinity of a Hopf bifurcation when choosing the driving parameters appropriately.


In conclusion, we have analysed the transition from Adler-type locking to Arnold-type locking and the emergence of a devil's staircase structure, in oscillators with modulated continuous forcing.  We have shown in theory and experiment that the periodic modulation of the driving force can introduce higher order locking tongues. The mechanism creating the devil's staircase is generic and the Arnold-type frequency locking appears in every nonlinear oscillator with periodically modulated forcing close to a Hopf bifurcation. Such oscillators show only a single main resonance tongue for a single driving frequency. With a modulated driving signal, we observe a complete devil's staircase in both the oscillation period of the amplitude and the average oscillation frequency. The harmonic resonances are induced by the nonlinear amplitude variations, and disappear when neglecting the amplitude dynamics. 
As a concrete example we studied the case of a periodically driven semiconductor laser, which conventionally shows only a single main resonance tongue when subject to single-frequency optical injection. With a modulated optical forcing, the semiconductor laser reproduces the Arnold-type locking structure very closely. 
Our results show that by proper implementation of an external drive, additional regions of stable devil's staircase frequency locking can be introduced in systems which, with constant driving, show only a single Adler-type resonance tongue. These higher order tongues may be of technological interest for applications where they could increase the density of injected optical combs in lasers, among other possibilities.
%

The authors acknowledge funding from Science Foundation Ireland under grant SFI 13/IA/1960.
BL acknowledges funding by the Deutsche Forschungsgemeinschaft (DFG, German Research Foundation) -- 404943123.

%

\end{document}